\documentclass[aps,showpacs,preprintnumbers,twocolumn,amsmath,amssymb,prl,floatfix]{revtex4} 

\usepackage{graphicx}

\usepackage{ifpdf}

\ifpdf
\usepackage{epstopdf}
\usepackage[pdftex]{hyperref}
\else
\usepackage[hypertex,ps2pdf,dvips,colorlinks,urlcolor=blue,citecolor=blue]{hyperref}
\fi

\newcommand\be{\begin{equation}}
\newcommand\bes{\begin{subequations}}
\newcommand\esu{\end{subequations}}
\newcommand\ee{\end{equation}}
\newcommand\erf[1]        {\eqref{#1}}
\newcommand{\ud}          {\mathrm d}
\newcommand\eps           {\varepsilon}
\newcommand\fii           {\varphi}
\newcommand\mc            {\mathcal}

\newcommand\p             {\partial}
\newcommand\psid          {\psi^{\dagger}}

\newcommand\kb            {k_\text{B}}
\newcommand\lam             {\lambda}
\newcommand\ga             {\gamma}
\newcommand\rhor          {\rho^{\text{(p)}}}
\newcommand\vev[1]{{\langle#1\rangle}}

\newcommand\doi[2]        {\href{http://dx.doi.org/#1}{#2}}

\newcommand\ibid          {{\it ibid.} }

\begin{document}

\title{Interaction quenches in the Lieb--Liniger model}
\author{M\'arton Kormos, Aditya Shashi, Yang-Zhi Chou, and Adilet Imambekov}
\affiliation{Department of Physics and Astronomy, Rice
University, Houston, Texas 77005, USA}
\date{\today}

\begin{abstract}

We obtain exact results on interaction quenches in the 1D Bose gas
described by the integrable Lieb-Liniger model.  We show that in the
long time limit integrability leads to significant deviations from the
predictions of the grand canonical ensemble and a description within
the generalized Gibbs ensemble (GGE) is needed. For a non-interacting
initial state and arbitrary final interactions, we find that the
presence of infinitely many conserved charges generates a non-analytic
behavior in the equilibrated density of quasimomenta. This manifests
itself in a dynamically generated Friedel-like oscillation of the
non-local correlation functions with interaction dependent oscillation
momenta. We also exactly evaluate local correlations and the
generalized chemical potentials within GGE.

\end{abstract}


\maketitle

Whether and how an isolated quantum system equilibrates or thermalizes
are fundamental questions in understanding non-equilibrium
dynamics. The answers can also shed light on the applicability of
quantum statistical mechanics to closed systems. While these questions
are very hard to study experimentally in the condensed matter setup,
they have become accessible in ultracold quantum gases due to the
recent experimental advances \cite{review}. Thanks to their
unprecendented tunability, ultracold atomic systems allow for the
study of non-equilibrium quantum dynamics of almost perfectly isolated
strongly correlated many-body systems in a controlled way. These
experiments \cite{exp, kinoshita} triggered a revival of
theoretical studies on issues of thermalization \cite{rev}.
Fundamental questions include whether stationary values of local
correlation functions are reached in a system brought out of
equilibrium, and if so, how they can be characterized. Can
conventional statistical ensembles describe the state? Is there any
kind of universality in the steady state and the way it is approached?

The absence of thermalization of a 1D bosonic gas reported in
\cite{kinoshita} brought to light the special role of
integrability. The lack of thermalization was attributed to the fact
that the system was very close to an integrable one, the Lieb--Liniger
(LL) model \cite{LL}, which is the subject of our Letter. The dynamics
of integrable systems is highly constrained by the presence of a large
number of so-called conserved charges in addition to the total particle
number, momentum, and energy. Thus integrable systems are not expected
to thermalize. Following the ideas of the subjective statistical
mechanics of Jaynes \cite{jaynes}, the so-called generalized Gibbs
ensemble (GGE) was proposed \cite{GGE} to capture the long-time
behavior of integrable systems brought out of equilibrium. This
ensemble is the least biased statistical representation of the system
once the conserved charges $\{Q_m\}$ are taken into account. The
density matrix is
\be \hat\rho_{\text{GGE}}=\frac{e^{-\sum_m \beta_m\hat Q_m}}{Z_\text{GGE}}\,,
\label{eq:GGErho}
\ee
where the generalized ``chemical'' potentials
$\{\beta_m\}$ are fixed by the expectation values $\vev{\hat Q_m}$, and
$Z_\text{GGE}=\mathrm{Tr}\left[e^{-\sum_m \beta_m\hat Q_m}\right]$.

The GGE was tested and its drawbacks were analyzed by various
numerical and analytical approaches \cite{gge_various}. However, with
a few exceptions \cite{cauxpapers,mossel}, only models that can be
mapped to quadratic bosonic or fermionic systems have been considered,
where the conserved charges are given by the mode occupation
numbers. While some of these models are paradigmatic, like the Ising
or Luttinger models, a prominent class of non-trivial integrable
systems has not been sufficiently explored, namely those solvable by
the Bethe Ansatz. In these models, the conserved charges are local
and cannot be expressed as mode occupations.

In this Letter, we derive experimentally testable predictions for the
long time behavior of the system after an interaction
quench~\cite{Muth, Gritsev} in the LL model by combining its Bethe
Ansatz solution and GGE. For a non-interacting initial state and
arbitrary final interactions, the infinitely many conserved charges
result in Friedel-like oscillations in the two-point functions, in
striking contrast to the grand canonical ensemble (GCE).

\paragraph*{The model.---} The LL model describes a system of identical
bosons in 1D interacting via a Dirac-delta potential.
The Hamiltonian in second quantized formulation is given by \cite{LL}
\be
\hat H= \int_0^L\ud x\,\left(\p_x\hat\psi^\dagger\p_x\hat\psi+
c\,\hat\psi^\dagger\hat\psi^\dagger\hat\psi\hat\psi\right)\,,
\label{eq:H_NLS}
\ee
where $c>0$ in the repulsive regime we wish to study, and 
for brevity we have set $\hbar=1$ and the boson mass to be equal to
$1/2$. The dimensionless coupling constant is given by
$\ga=c/n$, where $n=N/L$ is the density of the gas. In cold atom
experiments $\ga$ is a function of the 3D scattering length and the 1D
confinement \cite{olshanii,kinoshita}. The exact spectrum and thermodynamics of
the model can be obtained via Bethe Ansatz \cite{LL,KBI}.  Each
eigenstate of the system with $N$ particles on a ring of circumference
$L$ is characterized by a distinct set of quantum numbers $\{I_j\}$
that are integers (half-integers) for $N$ odd (even). The wave
function can be expressed in terms of $N$ quasimomenta $\{\lam_j\}$ that
satisfy a set of algebraic equations
\be
L\lam_j+\sum_{k=1}^N\theta(\lam_j-\lam_k) = 2\pi I_j\,,
\label{eq:BAE}
\ee
where $\theta(\lam)=2\arctan(\lam/c)$. The expectation values of the {\em
local} conserved charges can be computed as
\be
Q_m\equiv \vev{\hat Q_m}= \sum_j \lam_j^m\,,
\label{eq:Q}
\ee
in particular, the energy of a given state is simply
$E=Q_2=\sum_j\lam_j^2$. The wave function is identically zero if any two
of the $\{I_j\}$ coincide, which is reminiscent of the Pauli principle
for fermions. 
In the thermodynamic limit, a mixed state
corresponding to thermal equilibrium \cite{yang} is captured by a
filling fraction $0<f_I<1$ in the space of quantum numbers. It has
been shown in Ref.~\cite{mossel} that GGE also implies that such a
description is possible.
For calculations, it is more convenient to define a function $f(\lam)$
in terms of the quasimomenta. Unlike the case of free fermions,
where the relation between $f_I$ and $f(\lam)$ is trivial, here all
quasimomenta are coupled to each other by Eq.~\eqref{eq:BAE}.
Thus the density of occupied quasimomenta $\rhor(\lam)$ is not independent of
$f(\lam)$ but satisfies
the integral equation 
\cite{LL}
\begin{subequations}
\label{eq:rho}
\begin{align}
\rho(\lam)&=\frac1{2\pi} + \int\frac{\ud
  \lam'}{2\pi}\,\fii(\lam-\lam')\,\rhor(\lam')\,,\label{eq:rhoeq}\\
\rhor(\lam)&=f(\lam)\rho(\lam)
\end{align}
\end{subequations}
with the kernel $\fii(\lam)=2c/(\lam^2+c^2)$. Here $\rho(\lam)$ is
the maximal allowed density of quasimomenta and $\rhor(\lam)$ is
normalized as $\int\ud\lam\,\rhor(\lam)=n$.

\paragraph*{Conserved charges and the problem of moments.---}

The simplest way to bring a system out of equilibrium is a sudden
change of one of its parameters, a quantum quench.  In a cold atom
setting such a quench could be achieved by a rapid change of the
transverse confinement or the scattering length.  We will compute the
predictions of the GGE for a sudden quench of the interaction
parameter, and compare them to those of the GCE.  In order to describe
the final state in terms of $\rhor(\lam)$, one needs to find the
expectation values of the conserved charges $Q_m$ right after the
quench.  For $c>0,$ all solutions of Eq.~\erf{eq:BAE} are
real~\cite{KBI}, thus finding the density $\rhor(\lam)$ which
reproduces these charges is equivalent to solving the problem of
moments defined by \be Q_m=L\int\ud\lam\,\rhor(\lam)\,\lam^m\,.
\label{eq:Qrho}
\ee

The first few conserved charge operators $\hat Q_m$ can be written in
terms of the field operator $\hat\psi(x)$ as $\hat Q_0=\int\ud
x\,\hat\psi^\dagger\hat\psi,$ $\int\hat Q_1=-i\int\ud
x\,\hat\psi^\dagger\p_x\hat\psi,$ and $\hat Q_2=\hat H$ is the
Hamiltonian given by Eq.~\erf{eq:H_NLS}.  Except for $\hat Q_0$ and
$\hat Q_1$, the charges depend on $c$ and their expectation values
change during the quench according to their $c$-dependence
\cite{Muth}.  Unfortunately, the corresponding expressions are not
known explicitly for the operators $\hat Q_{m}$ for $m\ge4$
\cite{davies}.

In the thermodynamic limit, the even conserved charges are expected
to have the form 
\be
\hat Q_{2m}=\int\!\ud x\left[\{\text{deriv.}\}+
A_m \,c^m :\!\left(\hat\psi^\dagger\hat\psi\right)^{m+1}\!:\right]\,,
\label{eq:Jop}
\ee  
where $\{\text{deriv.}\}$ denotes unknown terms which involve products of
spatial derivatives of $\psi(x)$ and $\psid(x)$ and $:\,:$ denotes
normal ordering. The constants $A_m$ can be calculated comparing
Eq.~\erf{eq:Jop} to the semiclassical limit ($\gamma\ll1$)
where the expressions for the conserved charges are known
\cite{ZSgutkin}. The constants $A_m$ satisfy a certain recursion
relation, which can be solved to obtain
$A_m=2^m(2m-1)!!/(m+1)!$~\footnote{Alternatively, one can establish the
  values of $A_{m}$ by using the limiting semicircle behavior of the
  ground state solution in the weakly interacting
  regime~\cite{LL}, and evaluating $Q_m$ according to
  Eq~\erf{eq:Qrho}.}.
Note that assuming an even $\rhor(\lam)$, as in the
  ground state, we keep only the even charges.

\paragraph*{Universal semicircle solution.---}

\begin{figure}
\scalebox{0.33}{\includegraphics{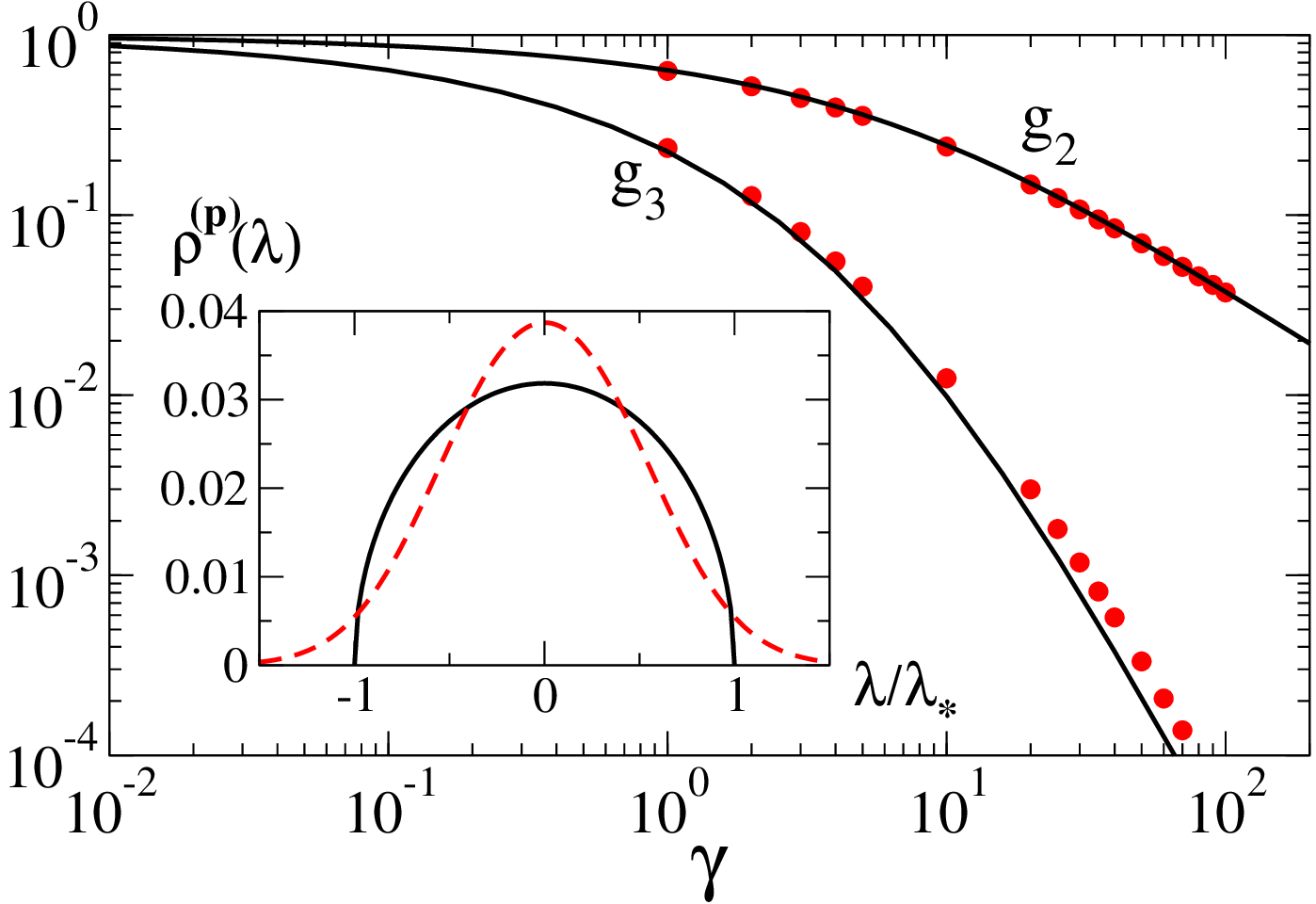}}
\caption{(Color online) Quench from a non-interacting initial state to
  arbitrary final interactions. Local correlations $g_2$ and $g_3$ as
  functions of the coupling $\ga$, calculated from the generalized
  Gibbs ensemble (solid, black) and from the grand canonical ensemble
  (dots, red). Inset: density of quasimomenta $\rhor(\lam)$ in the
  two ensembles for $\ga=100$.}
\label{fig:g2g3}
\end{figure}

If the initial state is in a pure non-interacting BEC (although we
expect our results to be also valid for small initial interactions),
then the expectation values of the unknown derivative terms in
Eq.~\erf{eq:Jop} vanish.
Thus the conserved charges are given by
$
Q_{2m}/L\approx
A_mc^m\vev{:\!(\hat\psi^\dagger\hat\psi)^{m+1}\!:}\approx
A_m n^{2m+1}\ga^m,$
where in the last step we used the fact that the local correlators
$\vev{\hat\psi^{\dagger m}\hat\psi^m}\to n^m$ as $\ga\to0$. 
We are left with the following problem of moments:
\be
\int\ud\lam \,\rhor(\lam)\,\lam^m=2^m\frac{(2m-1)!!}{(m+1)!}n^{2m+1}\ga^m\,,
\ee
which has the unique solution
\be
\rhor(\lam)= \frac1{\pi\sqrt\ga}\sqrt{1-\frac{\lam^2}{\lam^2_*}}\,,
\qquad \lam_*=2n\sqrt\ga
\label{eq:blown}
\ee
on the interval $[-\lam_*,\lam_*]$ and
zero otherwise (see inset of Fig.~\ref{fig:g2g3}). 
Using Eqs.~\erf{eq:rho} we can determine
$f(\lam)$: 
\begin{equation}
f(\lam) = 
\sqrt{\frac{2\left(1-\left(\lam/\lam_*\right)^2\right)}
  {\frac{4+\ga}4-\left(\lam/\lam_*\right)^2+
\sqrt{\left(\frac{4-\ga}4-\left(\lam/\lam_*\right)^2\right)^2+\ga}}}\,.
\label{eq:f}
\end{equation}
We emphasize that the existence of the sharp edge of support in
Eq.~\erf{eq:blown} is a direct consequence of our accounting for an
{\it infinite} number of conserved charges.  Keeping any finite number
within GGE would smear the sharp edge.

\paragraph{Correlation functions in the final state.---}
Knowing the filling fraction $f(\lam)$ allows us to calculate some of
the correlation functions. First we calculate local correlators
{\em exactly} using the results of Ref.~\cite{g3} which
give analytic expressions for the local two and three-point
correlators for arbitrary states that are captured by a continuous
$f(\lam)$.
We compute $g_2=\vev{:\!(\hat\psi^\dagger\hat\psi)^2\!:}/n^2$ and
$g_3=\vev{:\!(\hat\psi^\dagger\hat\psi)^3\!:}/n^3$ both for 
both for GGE and the GCE by using the appropriate
$f(\lam)$. In the latter only the energy and the particle
densities are fixed to be the same as for the GGE, which determines
the temperature and the chemical potential.  The results are shown in
Fig.~\ref{fig:g2g3}. For interactions not too strong, the GGE results
are very close to the values for GCE which slightly
underestimates $g_2$ and overestimates $g_3$. The deviation is bigger
for $g_3$ and grows with $\ga$. Both GGE and GCE results are
consistent with the outcomes of recent numerical simulation
\cite{Muth}. We conclude that the local correlators above are too simple to
effectively distinguish between the various ensembles.

\begin{figure}[t]
\centerline{\scalebox{0.33}{\includegraphics{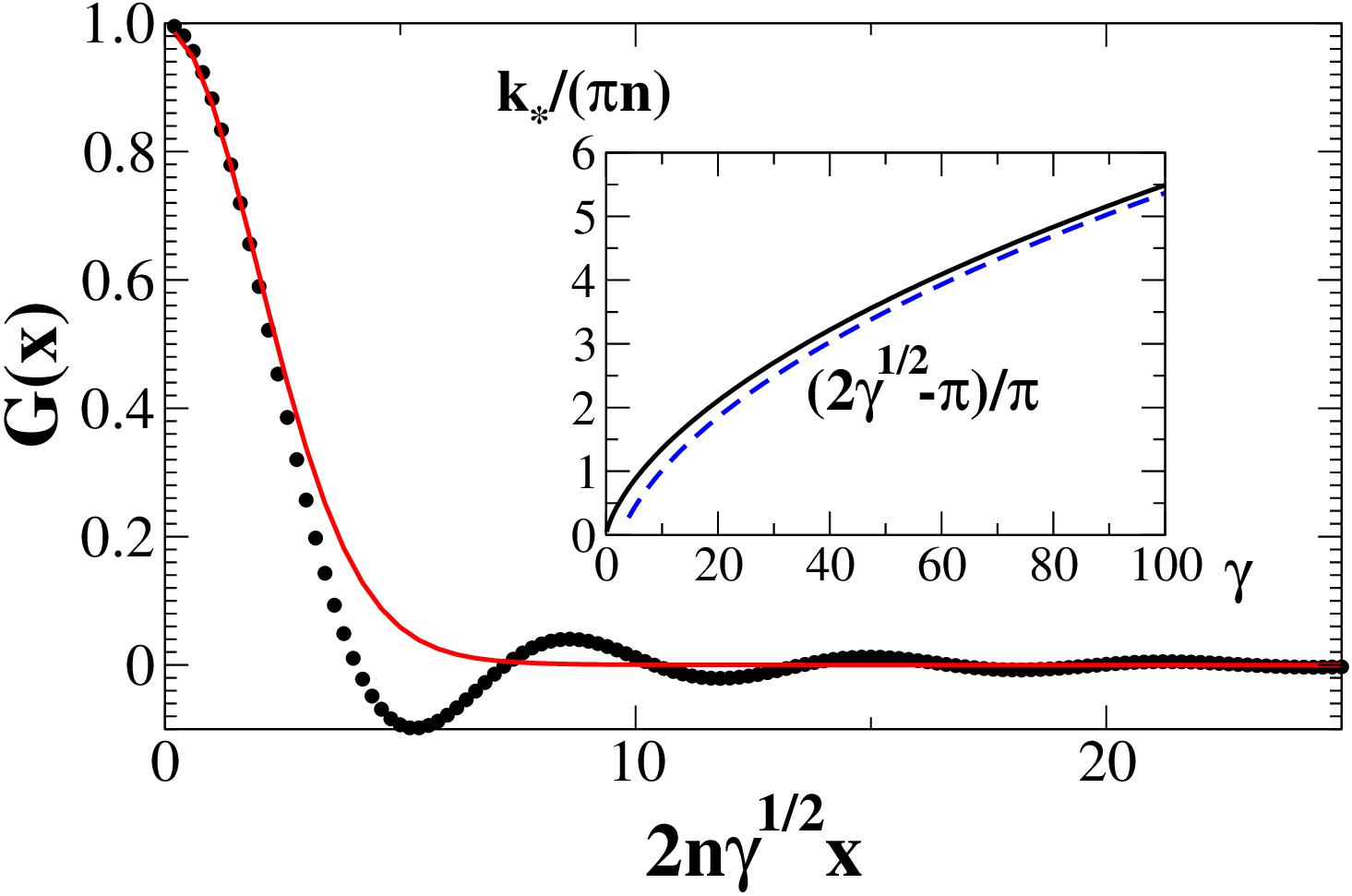}} }
\caption{(Color online) Equal time Green's function for quasimomenta
  distributed with the semi-circle function (dots) and with
  the thermal distribution (solid, red) when $\ga = 400$. Inset:
  dynamically generated Friedel-like oscillation momentum $k_*$ as a
  function of $\gamma$ in units of $k_\text{F}=\pi n$ (solid, black)
  with the large $\ga$ asymptotic result (dashed, blue).}
\label{fig:gx}
\end{figure}

Note, however, that the densities of filled states $\rhor(\lam)$ are
very different for the GGE and the GCE, see the inset of
Fig.~\ref{fig:g2g3}. Remarkably, the semicircle distribution exhibits
a non-analytic square root singularity at the edges, which has
immediate consequences for the behavior of the nonlocal correlation
functions.  In analogy with the Friedel oscillations, which appear due
to the existence of a sharp Fermi surface, we expect that the
non-analyticity of $\rhor(\lam)$ at $\lam_*$ will result in the
oscillating behavior of nonlocal correlation functions. Unlike the
period of conventional Friedel oscillations which depends only on the
average density, the period of these dynamically generated Friedel
oscillations also depends on the interaction strength. The
characteristic momentum $k_*$ which plays the role of the Fermi
momentum can be calculated using standard techniques~\cite{LL, KBI},
and is plotted in the inset of Fig.~\ref{fig:gx}. While we cannot
explicitly calculate the nonlocal correlation functions for arbitrary
final interactions, below we present detailed results for strong final
interactions which support this picture.

\paragraph*{Strongly interacting final state.---}

For large coupling the system is effectively in the fermionized
regime, since the largest quasimomentum $\lam_*$ scales only as $\sim\sqrt
c$ and thus the convolution term in Eq.~\erf{eq:rhoeq} depending on
the kernel $\fii(\lam)$ can be neglected. The function $f(\lam)$
takes the form 
\be 
f(\lam)\underset{\ga\to\infty}{\approx}\frac2{\sqrt\ga} 
\sqrt{1-\left(\frac\lam{\lam_*}\right)^2}\,. 
\label{eq:TG_SC}
\ee
Bosonic correlation functions can be
calculated by first fermionizing the field operators using Jordan--Wigner
strings,
$\hat\psi(x) = 
\exp[i\pi\int_{-\infty}^x\hat\psi^\dagger_\text{F}(z)
\hat\psi_\text{F}(z)dz]\hat\psi_\text{F}(x)$,
and then exploiting free fermionic correlators of
$\hat\psi_\text{F}$ in the large $\ga$ regime. Let us consider the equal time Green's function
\be
G(x) = \langle \hat\psi^\dagger(x)\hat\psi(0)\rangle_\text{SC}\,,
\ee
where SC denotes averaging over the semi-circle
distribution of Eq.~\erf{eq:TG_SC}. We proceed by introducing a
lattice discretization $a\ll 1/n, x=(m+1) a$ and recasting the
correlator as
\be
G(x) = 
\left\langle \!\psi^\dagger_F(x)
\prod_{k\leq m}\left[1-2\,\hat\psi^\dagger_F(ka)\hat\psi_F(ka)\right]
\hat\psi_F(a)\!\right\rangle\,.
\label{eq:GTGlat}
\ee
The long chain of operators is now amenable to a Wick expansion
using as a building block the {\em fermionic} two point function given
by the Fourier transform of the distribution \erf{eq:TG_SC}, $ G_{\rm
  FF}(x) =\int \frac{\ud\lam}{2\pi} f(\lam) e^{i\lam x} =
2n\,J_1(\lam_*x)/(\lam_*x), $ where $J_1(x)$ is the Bessel function of
the first kind. The Wick expansion of Eq.~\erf{eq:GTGlat} can be
recast as a Fredholm-like determinant since the correlator of $2n$
fields can be represented as a determinant of size $n$, and we need to
sum over all discretizations of the interval $(0,x)$ which can be
accomplished via a determinant expansion \cite{ripplezvon}. Finally,
we have $G(x) = \lim_{a\to
  0}\mathrm{det}_{m+1}(2a\mathbf{B}-\mathbf{A})/(2a)$ with
$A_{ij}=\text{diag}\{0,1,\dots,1\}$ and
$B_{ij}=G_\text{FF}\left((i-j)a\right)$ for $i,j>1$.  The limit $a\to
0$ is taken numerically.  For the semicircle distribution we find that
the correlation function can be well fitted using the Bessel function
with an exponentially decaying term,
$2n\,J_1(\lam_*x)/(\lam_*x)e^{-\alpha x}$. The value of $\alpha$
changes from $2.25$ to $2$ as $\ga$ is varied from $100$ to infinity.
A field theoretical understanding of the behavior of correlation
functions, possibly in the spirit of models going beyond the Luttinger
Liquid in 1D~\cite{RMPreview} remains to be achieved.

We can also compute the density-density correlation function
$
g_2(x)=\vev{\hat\psid(x)\hat\psid(0)\hat\psi(0)\hat\psi(x)}/n^2
$
for large final $\ga$ using the first few terms of the
infinite series given in Ref.~\cite{bogoliubov}. 
In the large $\ga$ limit the leading order for
arbitrary $f(\lam)$ is given by 
$
g_2(x)\approx 1-\left(\int\frac{\ud\lam}{2\pi}f(\lam)e^{i\lam x}\right)^2.
$ 
For the GGE case we use the $f(\lam)$ given in Eq.~\erf{eq:TG_SC}. The
result, $g_2(x)\approx 
1-2nJ_1^2(\lam_*x)/(\lam_*x)^2$, is oscillatory and decays as
a power law $\sim1/x^3$, in contrast to the exponential decay in the
thermal state.

\paragraph{GGE generalized chemical potentials.---}

\begin{figure}[t]
\centerline{\scalebox{0.33}{\includegraphics{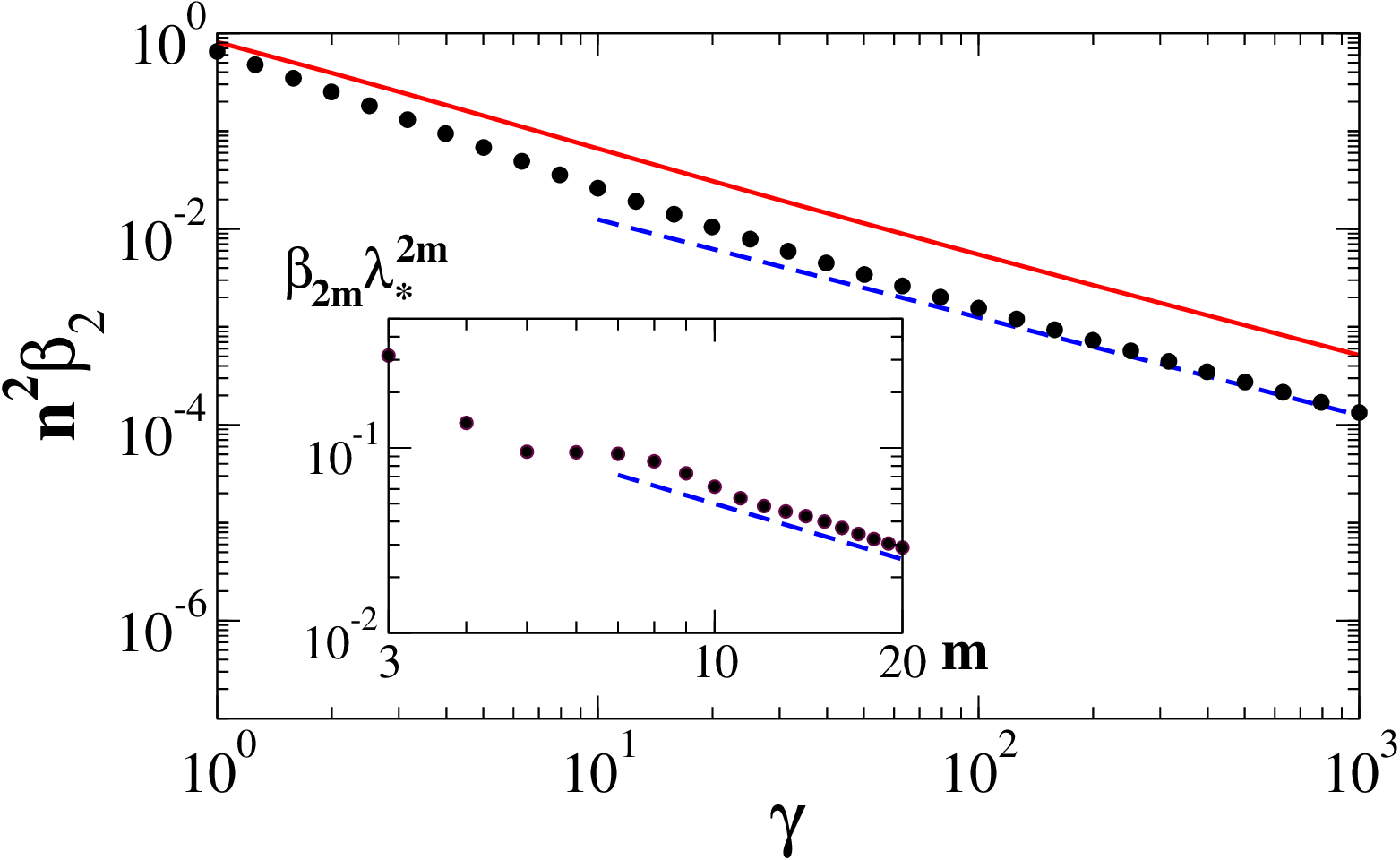}} }
\caption{(Color online) Dimensionless inverse temperature $n^2\beta_2$
  (dots) as a function of $\gamma,$ shown together with the grand
  canonical prediction (solid, red) and analytical $\gamma\gg 1$ result
  $\approx1/(8\gamma)$ (dashed, blue). Inset: generalized
  dimensionless chemical potential $\beta_{2m}\lam_*^{2m}$ as a
  function of $m$ for $\ga=1$ with large $m$ result \erf{eq:b2m}
  indicated.}
\label{fig:betag1}
\end{figure}

The GGE density matrix \erf{eq:GGErho} minimizes the generalized
free energy, $\mc{F}/(\kb T)=\sum_m\beta_m Q_m-S/\kb$. Expressing
$\{Q_m\}$ and the entropy $S$ in terms of $\rho(\lam)$ and $f(\lam)$, we can
obtain a nonlinear integral equation for the filling fraction
$f(\lam)=1/(e^{\eps(\lam)}+1)$ following the standard
procedure~\cite{yang,mossel,unpubl}:
\be
\eps(\lam) + \int\frac{\ud
  \lam'}{2\pi}\fii(\lam-\lam')\log\left(1+e^{-\eps(\lam')}\right)=
\sum_m\beta_m \lam^m \,.
\label{eq:TBA}
\ee
Here $\beta_2 \equiv \beta=1/(\kb T)$ is the inverse temperature.
Note that keeping only the particle number $Q_0=N$ and energy $Q_2=E$,
one recovers the usual GCE in Eq.~\erf{eq:GGErho}
and the standard Yang--Yang equation \cite{yang} in Eq.~\erf{eq:TBA}.

Since we know $\eps(\lam)=\log[1/f(\lam)-1]$ from the solution of the
problem of moments given by Eq.~\erf{eq:f}, the generalized chemical
potentials $\{\beta_{2m}\}$ can be determined by expanding the left
hand side of Eq.~\erf{eq:TBA} in powers of $\lam/\lam_*$. The
convolution term is a smooth bounded function, so for large $m$ its expansion
coefficients
decay exponentially fast. However, the coefficients of $\eps(\lam)$
decay only algebraically, since it diverges at $\lambda \to \pm
\lambda_*$ as $\eps(\lam)\sim -\frac12\log[1-(\lam/\lam_*)^2]$. This
divergence results in the large $m$ behavior of the chemical potentials as
\be
\beta_{2m}\approx \frac1{2m\lam_*^{2m}}=\frac1{2m(4n^2\ga)^m}\,.
\label{eq:b2m}
\ee
None of them diverges but since there are infinitely many of them,
their collective behavior renders $\eps(\lam)$ divergent for
$\lam\ge\lam_*$. This creates a ``Fermi edge'' by constraining
$\rhor(\lam)$ to have a finite range of support. This mechanism is markedly
different from the case of the ground state, where the divergence of a
{\it finite} number of chemical potentials, as the temperature goes to
zero, results in a step distribution function.

\paragraph{Summary.---}

We studied the large time behavior of the post interaction quench
state in the Lieb-Liniger model using analytic techniques by combining
the generalized Gibbs ensemble and Bethe Ansatz integrability of the
model. For a non-interacting initial state and arbitrary final
interactions, the infinitely many conserved charges result in
Friedel-like oscillations in the two-point functions with interaction
dependent momentum. This is in striking contrast to
the behavior of the thermalized system and provides a smoking gun
signal for cold atom experiments.  We also exactly evaluated local
correlations and the generalized chemical potentials within GGE.

We acknowledge funding from The Welch Foundation, Grant No. C-1739,
from the Sloan Foundation and from the NSF Career Award
No. DMR-1049082.

\end{document}